\def\d{\partial}
\def\dh{\mathop{\vphantom{\odot}\hbox{$\partial$}}}
\def\dl{\dh^\leftrightarrow}
\def\sqr#1#2{{\vcenter{\vbox{\hrule height.#2pt\hbox{\vrule width.#2pt
height#1pt \kern#1pt \vrule width.#2pt}\hrule height.#2pt}}}}
\def\w{\mathchoice\sqr45\sqr45\sqr{2.1}3\sqr{1.5}3\,}

\def\=d{\,{\buildrel\rm def\over =}\,}

\def\i3p{\p32\int d^3p}

\def\As{A\hbox to 1pt{\hss /}}
\def\np4{\int d^4p_1\cdots d^4p_{n-1}\, }

\def\nx4{\int d^4x_1\ldots d^4x_n\, }

\def\kon#1#2{\vbox{\halign{##&&##\cr
\lower4pt\hbox{$\scriptscriptstyle\vert$}\hrulefill &
\hrulefill\lower4pt\hbox{$\scriptscriptstyle\vert$}\cr $#1$&
$#2$\cr}}}

\def\konv#1#2#3{\hbox{\vrule height12pt depth-1pt}
\vbox{\hrule height12pt width#1cm depth-11.6pt}
\hbox{\vrule height6.5pt depth-0.5pt}
\vbox{\hrule height11pt width#2cm depth-10.6pt\kern5pt
      \hrule height6.5pt width#2cm depth-6.1pt}
\hbox{\vrule height12pt depth-1pt}
\vbox{\hrule height6.5pt width#3cm depth-6.1pt}
\hbox{\vrule height6.5pt depth-0.5pt}}
\def\konu#1#2#3{\hbox{\vrule height12pt depth-1pt}
\vbox{\hrule height1pt width#1cm depth-0.6pt}
\hbox{\vrule height12pt depth-6.5pt}
\vbox{\hrule height6pt width#2cm depth-5.6pt\kern5pt
      \hrule height1pt width#2cm depth-0.6pt}
\hbox{\vrule height12pt depth-6.5pt}
\vbox{\hrule height1pt width#3cm depth-0.6pt}
\hbox{\vrule height12pt depth-1pt}}

\def\konw#1#2#3{\hbox{\vrule height12pt depth-1pt}
\vbox{\hrule height12pt width#1cm depth-11.6pt}
\hbox{\vrule height6.5pt depth-0.5pt}
\vbox{\hrule height12pt width#2cm depth-11.6pt \kern5pt
      \hrule height6.5pt width#2cm depth-6.1pt}
\hbox{\vrule height6.5pt depth-0.5pt}
\vbox{\hrule height12pt width#3cm depth-11.6pt}
\hbox{\vrule height12pt depth-1pt}}

\def\i{{\rm int}}

\def\m3{{\mu_1\mu_2\mu_3}}

\def\p{{(+)}}

\hsize=14 cm \vsize=20.8 cm \tolerance=400
\hoffset=2cm
\font\trm = cmr10 scaled \magstep3
\font\srm = cmr10 scaled \magstep2

\voffset=2cm
\scriptscriptfont0 =\scriptfont0
\scriptscriptfont1 =\scriptfont1

\nopagenumbers
\vbox to 1.5cm{ }
\centerline{Preprint: ZU-TH-9/96, hep-th/9606100}
\vbox to 1.5cm{ }
\centerline{\trm Non-Uniqueness of Quantized Yang-Mills Theories }
\vskip 2cm
\centerline{\srm Michael D\"utsch } \vskip 0.5cm
\centerline{\it
Institut f\"ur Theoretische Physik der Universit\"at Z\"urich}
\centerline{\it Winterthurerstr. 190, CH-8057 Z\"urich, Switzerland}\vskip 3cm
{\bf Abstract.} -We consider quantized Yang-Mills theories in the framework of
causal perturbation
theory which goes back to Epstein and Glaser.
In this approach gauge invariance is expressed by a simple commutator relation
for the
S-matrix. The most general coupling which is gauge invariant in first order
contains a
two-parametric ambiguity in the ghost sector - a divergence- and a
coboundary-coupling may
be added. We prove (not completely)
that the higher orders with these two additional couplings
are gauge
invariant, too. Moreover we show that the ambiguities of the $n$-point
distributions
restricted to the physical subspace are only a sum of divergences (in the sense
of vector analysis).
It turns out that the theory without divergence- and coboundary-coupling is the
most
simple one in a quite technical sense.
The proofs for the $n$-point distributions containing coboundary-couplings are
given
up to third or fourth order only, whereas the statements about the
divergence-coupling are proven in
all orders.
chromodynamics.
\vfill\eject
\pageno=1
\headline={\tenrm\ifodd\pageno\hss\folio\else\folio\hss\fi}
\footline={\hss}
{\trm 1. Introduction}
\vskip 1cm
{\it 1.1 The Model}
\vskip 0.5cm
In a recent series of papers [1-5] non-abelian gauge invariance has been
studied
in the framework of causal perturbation theory [6,7]. This approach, which goes
back to
Epstein and Glaser [6], has the merit that one works exclusively with free
fields, which
are mathematically well-defined, and performs only justified operations with
them.

In causal perturbation theory one makes
an ansatz for the S-matrix as a formal power series in the coupling constant
$$S(g_0,g_1,...,g_l)=1+\sum_{n=1}^{\infty}{1\over n!}\sum_{i_1,...,i_n=0}^l\int
d^4x_1...d^4x_n\,
T_n^{i_1...i_n}(x_1,...,x_n)g_{i_1}(x_1)...g_{i_n}(x_n).\eqno(1.1)$$
The indices $i\in\{0,1,...,l\}$ label different couplings $T^i_1$, which are
switched by
different test functions $g_i\in{\cal S}({\bf R}^4)$. The operator-valued
distribution
$T_n^{i_1...i_n}(x_1,...,x_n)$ has a vertex of the type $T^{i_s}_1$ at
$x_s\>(1\leq s\leq n)$.
The $T_n$'s are constructed
inductively from the given first order (see appendix A). In our model the
$i=0$-coupling
$$T_1^0(x)\=d T_1^{0A}(x)+T_1^{0u}(x),\eqno(1.2)$$
is the usual three-gluon coupling
$$T^{0A}_1(x)\=d {ig\over 2}f_{abc}:A_{\mu a}(x)A_{\nu
b}(x)F^{\nu\mu}_c(x):,\eqno(1.3)$$
plus the usual ghost coupling
$$T^{0u}_1(x)\=d -igf_{abc}:A_{\mu a}(x)u_b(x)\d^\mu\tilde u_c(x):.\eqno(1.4)$$
Herein $g$ is the coupling constant and $f_{abc}$ are the structure constants
of the group
SU(N). The gauge potentials $A^\mu_a,\>F_a^{\mu\nu}\=d \d^\mu A_a^\nu -\d^\nu
A_a^\mu $,
and the ghost fields $u_a,\>\tilde u_a$ are massless and fulfil the wave
equation.
(We work throughout in the Feynman gauge $\lambda=1$.)

{\it Gauge invariance} means roughly speaking
that the commutator of the $T_n^{0...0}$-distribu\-tions with the gauge charge
$$Q\=d \int_{t=const.} d^3x\,(\d_{\nu}A_a^{\nu}{\dl}_0u_a)\eqno(1.5)$$
is a (sum of) divergence(s) (in the sense of vector analysis). In first order
this holds true
$$[Q,T^0_1(x)]=i\d_\nu T^{1\nu}_1(x),\eqno(1.6)$$
where
$$T^{1\nu}_1(x)\=d igf_{abc}[:A_{\mu a}(x)u_b(x)F_c^{\nu \mu}(x):
-{1\over 2}:u_a(x)u_b(x)\d^\nu\tilde u_c(x):].\eqno(1.7)$$
We choose this expression to be the $i=1$-coupling in (1.1) and call it a
Q-vertex.
Note that $[Q,T^{0A}_1]$ alone is not a divergence. In order to have gauge
invariance in first
order, we are forced to introduce the ghost coupling $T^{0u}_1$ (1.4). However,
the latter
coupling is not uniquely fixed by this procedure. The present paper deals with
these
ambiguities. We define gauge
invariance in arbitrary order [2] by
$$[Q,T_n^{0...0}(x_1,...,x_n)]=i\sum_{l=1}^n\d_\nu^{x_l}T^{0...010...0\nu}_n(x_1,...,x_n),\eqno(1.8)$$
where the upper index $1$ in $T^{0...010...0}_n$ is at the $l$-th position.
The divergences on the r.h.s. of (1.8) are precisely specified:
$T^{0...010...0}_n(x_1,...,x_n)$
is the $T_n$-distribution of (1.1) which has a Q-vertex (1.7) at $x_l$ and all
other vertices are
$T_1^0$-couplings (1.2). Gauge invariance (1.8), which has been proven in all
orders $n$
[1-5], implies the invariance of the S-matrix $S(g,0,...,0)$ (1.1) with respect
to simple gauge
transformations of the {\it free} fields [5]. These transformations are the
{\it free field
version of the famous BRS-transformations} [8]. Moreover {\it unitarity on the
physical
subspace} [4] can be proven by means of gauge
invariance (1.8). The C-number identities expressing (1.8) imply the
{\it Slavnov-Taylor identities} [9]. Finally we mention that the four-gluon
interaction is a
normalization term in second order, which is uniquely fixed by gauge invariance
(see [1,5] and (2.59)).

Let us turn to the mentioned non-uniqueness in the ghost sector. The most
popular method to derive the
ghost-coupling is the one of Faddeev and Popov. However, this method of
quantization contains
loopholes (even in perturbation theory) [10]. Therefore, Beaulieu [10]
determined the quantum
Lagrangian from the requirement of its full BRS-invariance. We proceed in an
analogous way.
{\it Our aim is to work out the
most general Yang-Mills theory which is gauge invariant (1.8) in all orders and
to investigate
the physical and technical implications of the ambiguities.}
\vskip 0.5cm
{\it 1.2 Most General Coupling which is Gauge Invariant in First Order}
\vskip 0.5cm
In order to simplify the notations we define an operator $d_Q$ by means of our
gauge charge $Q$ (1.5)
$$d_QA\=d QA-(-1)^{Q_g}A(-1)^{Q_g}Q,\eqno(1.9)$$
where $Q_g$ is the ghost charge operator [11,12]
$$Q_g\=d i\int_{t=const.}d^3x:\tilde u_a(x){\dl}_0 u_a(x):,\quad\quad\quad
[Q_g,u_a]=-u_a,\quad [Q_g,\tilde u_a]=\tilde u_a.\eqno(1.10)$$
and $A$ is a suitable operator on the Fock space such that (1.9) makes sense.
If the ghost
charge of $A$ is an integer, $[Q_g,A]=zA,\>z\in {\bf Z}$, the expression (1.9)
is the
commutator or anticommutator of $Q$ with $A$. Note the product rule
$$d_Q(AB)=(d_QA)B+(-1)^{Q_g}A(-1)^{Q_g}d_QB.\eqno(1.11)$$
One easily verifies [1]
$$Q^2=0,\eqno(1.12)$$
which implies
$$(d_Q)^2=0.\eqno(1.13)$$
Because $d_Q$ is nilpotent, it can be interpreted as coboundary-operator
in the framework of a homological algebra [11]. (The gradiation is given by the
ghost charge (1.10).)
Therefore, we call an element of the range (kernel) of $d_Q$ a coboundary
(cocycle).

Let us add a coboundary
$$\beta_1d_QK_1(x),\quad\quad\quad\beta_1\in{\bf R}\>\>{\rm
arbitrary},\eqno(1.14)$$
with
$$K_1(x)\=d gf_{abc}:u_a(x)\tilde u_b(x)\tilde u_c(x):,\eqno(1.15)$$
to $T_1^0(x)$. Due to (1.13), gauge invariance in first order (1.6) remains
true with the same
Q-vertex $T_1^{1\nu}$ (1.7). Moreover, we add a divergence
$$\beta_2\d_\mu K_2^\mu(x),\quad\quad\quad\beta_2\in{\bf R}\>\>{\rm
arbitrary},\eqno(1.16)$$
with
$$K_2^\mu(x)\=d igf_{abc}:A^\mu_a(x)u_b(x)\tilde u_c(x):,\eqno(1.17)$$
to $T_1^0(x)$. Adding simultaneously $\beta_2d_QK_2^\nu(x)$ to $T_1^{1\nu}(x)$,
our gauge invariance (1.6) is obviously preserved. Are there further couplings
which
are gauge invariant in first order? The answer is 'no' [11,13], if the
following,
physically reasonable requirements are additionally imposed:

(A) The coupling is a combination of at least three free field operators.

(B) The coupling has mass-dimension $\leq 4$. This guarantees
(re)normalizability of the theory,
if the fundamental (anti)commutators have singular order $\omega ([A_a^\mu
,A_b^\nu)])=-2$ and
$\omega (\{u_a ,\tilde u_b)\})=-2$ (see appendix A and [2]).

(C) Lorentz covariance.

(D) SU(N)-invariance.

(E) The coupling has ghost charge zero: $[Q_g,T_1^0]=0$.

(F) Invariance with respect to the discrete symmetry transformations P,T and
C.

(G) Pseudo-unitarity $S_1(g_0^*,0,...,0)^K=S_1(g_0,0,...,0)^{-1}$ forces
$\beta_1,\,\beta_2$
to be real. ($S_1$ is the first order $n=1$ of (1.1) and
K is a conjugation which is related to the adjoint [4,12].)

Remarks: (1) The self-interaction of the gauge bosons $T_1^A$ (1.3) is unique.
There is
only an ambiguity in the ghost coupling.

(2) In [5] the coupling to fermionic matter fields in the fundamental
representation was
studied in detail. It is easy to see that the above requirements fix this
coupling uniquely.
Therefore, we do not consider matter fields in this paper.

\vskip 0.5cm
{\it 1.3 Outline of the Paper}
\vskip 0.5cm
The paper yields the following results:

(A) The higher orders with divergence- or coboundary-coupling (1.14-17) are
gauge invariant
for all values of $\beta_1,\beta_2\in{\bf R}$ (sects.2.2, 2.4). (For the
coboundary-coupling
this will be proven up to third order only.) The analogous result for the full
BRS-symmetry
in the usual Lagrangian approach is known in the literature, see e.g. [10].
However, only a
one-parametric ambiguity is studied there. This difference will be discussed in
sect.2.7, remark (4).

(B) We will show that the $T_n$'s with divergence-coupling are divergences with
respect to their
divergence-vertices (sect.2.2). The $T_n$'s $(1\leq n\leq 4)$ with
coboundary-coupling
are divergences too,
if they are restricted to the physical subspace [4] (sect.2.8). This will be an
immediate
consequence of a representation of these $T_n$'s, which will be proven in
sect.2.4.

(C) The results in higher orders about the divergence-coupling and partly the
results about the
coboundary-coupling are independent on the explicit expressions (1.2-4) and
(1.14-17)
of the couplings (sect.2.5). They apply to any gauge invariant quantum field
theory.

(D) Gauge invariance for second order tree diagrams requires normalization
terms, namely the usual
four-gluon interaction and a four-ghost interaction (sect.2.7). (The latter
appears only for
$(\beta_1,\beta_2)\not= (0,0)$.) By studying these normalization terms we will
find a
criterion which reduces the freedom in the choice of $\beta_1,\beta_2\in{\bf
R}$ to
a one-parametric set (sects.2.7-8). We will mention a second, quite technical
criterion
which gives another restriction of $\beta_1,\beta_2$ (sect.2.8).
Together we will see that the theory with $\beta_1=0=\beta_2$ is the most
simple one.

(E) The Q-vertex is not uniquely fixed by gauge invariance in first order
(1.6).
In order to prove gauge invariance in {\it higher} orders of the theory
$(T_1^0+\beta_1d_QK_1+\beta_2\d_\mu K_2^\mu),\>\beta_1,\beta_2\in{\bf R}$
((1.2-4),
(1.14), (1.16)), it is not necessary to modify the above introduced Q-vertex
((1.7) plus
$\beta_2d_QK_2^\nu$). Therefore, the ambiguity of the Q-vertex is not very
interesting.
Nevertheless we show in sect.2.3 that the possible modifications of the
Q-vertex do not destroy
gauge invariance in higher orders.

(F) In appendix (C) we assume certain identities to hold true. They concern
exclusively the
starting-coupling $T_1^0$ (1.2-4), its Q-vertex $T_1^1$ (1.7) and its
'Q-Q-vertex' $T_1^5$
introduced below (2.5), and are a kind of generalization of gauge invariance
(1.8).
A special case of this assumption is verified in appendix (B). By means of
these identities
we will be able to prove the results about the coboundary-coupling in {\it all}
orders.
\vskip 1cm
{\trm 2. Divergence- and Coboundary-Coupling in\break\vskip 0.2cm Higher
Orders}
\vskip 1cm
{\it 2.1 Preparations}
\vskip 0.5cm
In order to study the $T_n$'s with a divergence- (1.16) and/or a
coboundary-coupling (1.14) in
higher orders $n\geq 2$, we define a big theory which contains these couplings
and some
auxiliary vertices
$$S_1(g_0,g_1,...,g_7)\=d \int d^4
x\{T_1^0(x)g_0(x)+T_1^{1\nu}(x)g_{1\nu}(x)+T_1^2(x)g_2(x)
+T_1^{3\nu}(x)g_{3\nu}(x)+$$
$$+T_1^{4\nu}(x)g_{4\nu}(x)+T_1^{5\nu\mu}(x)g_{5\nu\mu}(x)+
T_1^6(x)g_6(x)+T_1^7(x)g_7(x)\},\eqno (2.1)$$
where $T_1^0,\,T_1^{1\nu}$ are given by (1.2-4) and (1.7), furthermore
$$T_1^{4\nu}(x)\=d\beta_2 K_2^\nu (x),\eqno(2.2)$$
$$T_1^2(x)\=d \d_\nu T_1^{4\nu}(x)=\beta_2 \d_\nu K_2^\nu (x),\eqno(2.3)$$
$$iT_1^{3\nu}(x)\=d d_QT_1^{4\nu}(x)=\beta_2 d_QK_2^\nu (x),\eqno(2.4)$$
$$T_1^{5\nu\mu}(x)\=d {igf_{abc}\over
2}:u_a(x)u_b(x)F^{\nu\mu}_c(x):=-T_1^{5\mu\nu}(x),\eqno(2.5)$$
$$T_1^6(x)\=d \beta_1 K_1(x),\eqno(2.6)$$
and
$$T_1^7(x)\=d d_Q T_1^6(x)=\beta_1 d_QK_1(x).\eqno(2.7)$$
For technical reasons the divergence-coupling $T_1^2$ (2.3) and the
coboundary-coupling
$T_1^7$ (2.7) are not directly added to $T_1^0$ - they both are smeared out
with a separate
test function.
The appearance of the vertex $T_1^{5\nu\mu}$ is motivated by the relation
$$d_QT_1^{1\nu}(x)=i\d_\mu
T_1^{5\nu\mu}(x).\eqno(2.8)$$
Therefore, we sometimes call $T_1^5$ 'Q-Q- vertex'. Furthermore note that
$T_1^{5\nu\mu}$ is a cocycle
$$d_QT_1^{5\nu\mu}(x)=0.\eqno(2.9)$$
The vertices $T_1^{1\nu},\>T_1^{3\nu}$ and $T_1^6$ are fermionic, all other
vertices
are bosonic. The first ones give rise to some additional minus-signs in the
inductive
construction of the $T_n$'s, but there is no serious complication (see the
appendix of [3]).
We are interested in the physically relevant theory
$$T_n(x_1,...,x_n)\=d
\sum_{i_1,...,i_n\in\{0,2,7\}}T_n^{i_1...i_n}(x_1,...,x_n),\eqno(2.10)$$
which corresponds to the choice $g\=d g_0=g_2=g_7\not= 0$ and
$g_1=0,\>g_{3\nu}=0,\>g_{4\nu}=0,\>
g_{5\nu\mu}=0$ and $g_6=0$ in the $n$-th order S-matrix $S_n(g_0,g_1,...,g_7)$.
Gauge
invariance in the sense (1.8) of this theory is formulated in terms of the
Q-vertices $T_1^{1\nu},\>
T_1^{3\nu}$ and $T_1^{8\nu}\=d 0$. This means in first order
$$d_Q T_1^0=i\d_\nu T_1^{1\nu},\eqno(2.11)$$
$$d_Q T_1^2=i\d_\nu T_1^{3\nu},\eqno(2.12)$$
$$d_Q T_1^7=0,\eqno(2.13)$$
and in arbitrary order $n$
$$d_Q T_n^{i_1...i_n}=i\sum_{l=1}^n\d_\nu^l
T_n^{i_1...i_{l-1}\,i_l+1\,i_{l+1}...i_n\>\nu},\eqno(2.14)$$
where $i_1,...,i_n\in\{0,2,7\}$ and
$$T_n^{i_1...8...i_n\>\nu}\=d 0.\eqno(2.15)$$
We shall often use that $T_n^{0...0}$ is gauge invariant (1.8) [1-5].
\vskip 0.5cm
{\it 2.2 Higher Orders with Divergence-Coupling}
\vskip 0.5cm
We are going to prove

{\bf Proposition 1}: {\it Choosing suitable normalizations, the relations}
$$F_n^{2...20...0}(x_1,...,x_n)=\d_{\mu_1}^1...\d_{\mu_r}^rF_n^{4...40...0\mu_1...\mu_r}(x_1,...,x_n),
\eqno(2.16)$$
$$F_n^{32...20...0\nu}(x_1,...,x_n)=\d_{\mu_2}^2...\d_{\mu_r}^rF_n^{34...40...0\nu\mu_2...\mu_r}
(x_1,...,x_n),\eqno(2.17)$$
$$F_n^{2...210...0\nu}(x_1,...,x_n)=\d_{\mu_1}^1...\d_{\mu_r}^rF_n^{4...410...0\mu_1...\mu_r\nu}
(x_1,...,x_n)\eqno(2.18)$$
{\it hold true for all $F=A',R',R'',D,A,R,T',T,\tilde T$ and in all orders
$n$.}

Remarks: (1) The assertions (2.16-18) are generalizations of (2.3) to arbitrary
orders and mean that
the divergence-structure of $T_1^2$ can be maintained by constructing the
higher orders.

(2) Due to the symmetrization (A.14) the $T_n^{...},\>\tilde T_n^{...}$ fulfil
$$T_n^{i_1...i_n}(x_1,...,x_n)=(-1)^{f(\pi )}T_n^{i_{\pi 1}...i_{\pi n}}(x_{\pi
1},...,x_{\pi n}),
\>\>\>\>\>\>\forall \pi \in {\cal S}_n,\eqno(2.19)$$
where the Lorentz indices are permuted, too, and $f(\pi )$ is the number of
transpositions of
fermionic vertices in $\pi$. Therefore, the equations (2.16-18) remain true for
$T_n,\,\tilde T_n$,
if the indices are permuted according to (2.19).

(3) We will see in the proof that the $T_n^{...4...}$'s on the r.h.s. can be
normalized in an
arbitrary symmetrical way. (A normalization is said to be symmetrical if the
corresponding
$T_n^{...}$ satisfies (2.19).) But the normalization of the $T_n^{...2...}$'s
on the l.h.s.
is uniquely fixed by the normalization of the $T_n^{...4...}$'s.

{\it Proof}: We show that (2.16-18) can be maintained in the inductive step
$(n-1)\rightarrow n$
described in appendix A. Obviously there are only two operations in this step
which need an
investigation, namely (A) the construction of the tensor products in
$A'_n,R'_n,R''_n$ (A.1-3) and
(B) the distribution splitting $D_n=R_n-A_n$ (A.7).

(A) Let us consider (2.17) for $A_n^{\prime ...}$ (A.2)
$$A_n^{\prime 32...20...0\nu}(x_1,...;x_n)=\sum_{X,Y,(x_1\in X)}\tilde
T_k^{32...20...0\nu}(X)
T_{n-k}^{2...20...0}(Y,x_n)+$$
$$+\sum_{X,Y,(x_1\in Y)}\tilde
T_k^{2...20...0}(X)T_{n-k}^{32...20...0\nu}(Y,x_n).\eqno(2.20)$$
Inserting the induction hypothesis (2.16-17) in lower orders $k,\>n-k$, we
obtain
$$(2.20)=\sum_{(x_1\in X)} \d_{\mu_2}^2...\d_{\mu_s}^s\tilde
T_k^{34...40...0\nu\mu_2...\mu_s}(X)
\d_{\mu_{s+1}}^{1}...\d_{\mu_r}^{r-s}T_{n-k}^{4...40...0\mu_{s+1}...\mu_r}(Y,x_n)+$$
$$+\sum_{(x_1\in Y)} \d_{\mu_1}^1...\d_{\mu_s}^s\tilde
T_k^{4...40...0\mu_1...\mu_s}(X)
\d_{\mu_{s+2}}^{2}...\d_{\mu_r}^{r-s}T_{n-k}^{34...40...0\nu\mu_{s+2}...\mu_r}(Y,x_n)=$$
$$=\d_{\mu_2}^2...\d_{\mu_r}^rA_n^{\prime
34...40...0\nu\mu_2...\mu_r}(x_1,...,x_n).\eqno(2.21)$$
The other verfications of (2.16-18) for $A'_n,\>R'_n,\>R''_n$ are completely
analogous.

(B) According to (A) the $D_n$'s (A.4) fulfil (2.16-18). Let
$R_n^{34...40...0\nu\mu_2...\mu_r}$
be an arbitrary splitting solution of $D_n^{34...40...0\nu\mu_2...\mu_r}$. Then
the definition
$$R_n^{32...20...0\nu}(x_1,...,x_n)\=d\d_{\mu_2}^2...\d_{\mu_r}^rR_n^{34...40...0\nu\mu_2...\mu_r}
(x_1,...,x_n),\eqno(2.22)$$
yields a splitting solution of $D_n^{32...20...0\nu}$, because
$R_n^{32...20...0\nu}$ (2.22)
has its support in $\Gamma^+_{n-1}(x_n)$ (A.6) and
$R_n^{32...20...0\nu}=D_n^{32...20...0\nu}$
on $\Gamma^+_{n-1}(x_n)\setminus \{(x_n,...,x_n)\}$. The procedure for (2.16),
(2.18) is
similar. $\quad\quad\w$

Applying $d_Q$ to (2.16) we see that $d_QT_n^{2...20...0}$ is a divergence
$$d_QT_n^{2...20...0}(x_1,...,x_n)=\d_{\mu_1}^1...\d_{\mu_r}^rd_QT_n^{4...40...0\mu_1...\mu_r}
(x_1,...,x_n),\eqno(2.23)$$
if there is at least one divergence-vertex $T_1^2$. However, the divergences on
the r.h.s. of
(2.23) are derivatives with respect to the divergence-vertices and generally
not to the Q-vertices. Consequently, (2.23) means not gauge invariance of
$T_n^{2...20...0}$
in the sense of (1.8) rsp. (2.14). In order to obtain the latter we will prove

{\bf Proposition 2}: {\it Starting with arbitrary symmetrical normalizations of
$T_n^{4...40...0}$
and $T_n^{4...410...0},...,T_n^{4...40...01}$, there exists a symmetrical
normalization of
$T_n^{34...40...0},...,T_n^{4...430...0}$ such that the equation}
$$d_QT_n^{4...40...0\mu_1...\mu_r}=i[T_n^{34...40...0\mu_1...\mu_r}+...+T_n^{4...430...0
\mu_1...\mu_r}+$$
$$+\d_\nu^{r+1}T_n^{4...410...0\mu_1...\mu_r\nu}+...+\d_\nu^nT_n^{4...40...01
\mu_1...\mu_r\nu}]\eqno(2.24)$$
{\it holds true in all orders $n$ and for $r=1,2,...,n$ vertices $T_1^4$ rsp.
$T_1^3$}.

Remarks: (1) The assertion (2.24) is a kind of gauge invariance equation, which
is a generalization
of (2.4) and (2.11) to higher orders.

(2) We will prove (2.24) for all $F_n,\>F=A',R',R'',D,A,R,T',T,\tilde T$ by
induction on $n$.

(3) Applying $\d_{\mu_1}^1...\d_{\mu_r}^r$ to (2.24) we obtain by means of
proposition 1

{\bf Corollary 3}: {\it With the normalization of (2.16) the distributions
$F_n^{2...20...0},
\>F=A',R',R'',D,A,R,T',T,\tilde T$ are gauge invariant, i.e. they fulfil
(2.14).}

{\it Proof of proposition 2}: The proof follows the inductive construction of
the $T_n$'s. Since
(2.24) is a linear equation, we merely have to consider the same operations (A)
(construction
of tensor products) and (B) (distribution splitting) as in the proof of
prop.1.

(A) Inserting the induction hypothesis (2.24) or gauge invariance (1.8) into
$d_Q\tilde T_k^{4...40...0}$ and $d_QT_{n-k}^{4...40...0}$ in
$$d_QA_n^{\prime 4...40...0\mu_1...\mu_r}(x_1,...;x_n)=\sum_{X,Y}[(d_Q\tilde
T_k^
{4...40...0\mu_1...\mu_s}(X))T_{n-k}^{4...40...0\mu_{s+1}...\mu_r}(Y,x_n)+$$
$$+\tilde
T_k^{4...40...0\mu_1...\mu_s}(X)d_QT_{n-k}^{4...40...0\mu_{s+1}...\mu_r}(Y,x_n)],\eqno(2.25)$$
one easily obtains that the $A'_n$-distributions fulfil (2.24), and similarly
this holds true for $R'_n,\,R''_n$.

(B) Let
$R_n^{4...40...0},\>R_n^{4...410...0},...,R_n^{4...40...01},\>R_n^{434...40...0},...,
R_n^{4...430...0}$ be arbitrary splitting solutions of
$D_n^{4...40...0},\>D_n^{4...410...0},...,
D_n^{4...40...01},\>D_n^{434...40...0},...,D_n^{4...430...0}$. Due to (A) the
$D_n$-distributions
fulfil (2.24). Since the operators $d_Q$ and $\d_\nu^s$ do not enlarge the
support of the
distribution to which they are applied, we obtain by the definition
$$iR_n^{34...40...0\mu_1...\mu_r}\=d d_QR_n^{4...40...0\mu_1...\mu_r}-$$
$$-i[R_n^{434...40...0\mu_1...\mu_r}+...+R_n^{4...430...0\mu_1...\mu_r}
+\d_\nu^{r+1}R_n^{4...410...0\mu_1...\mu_r\nu}+...+\d_\nu^nR_n^{4...40...01
\mu_1...\mu_r\nu}]\eqno(2.26)$$
a splitting solution of $iD_n^{34...40...0\mu_1...\mu_r}$. Obviously the
$T'_n\=d R_n-R'_n$-distributions fulfil (2.24) and this equation is maintained
in the symmetrization
$T'_n\rightarrow T_n$ (A.14). $\quad\quad\w$
\vskip 0.5cm
{\it 2.3 Non-Uniqueness of the Q-Vertex $T_1^{1\nu}$ in Higher Orders}
\vskip 0.5cm
The total Q-vertex $T_{1/1}^\nu\=d T_1^{1\nu}+T_1^{3\nu}$ of the theory (2.10)
is not
uniquely fixed by gauge invariance in first order
$d_Q(T_1^0+T_1^2+T_1^7)=i\d_\nu T_{1/1}^\nu$.
One has the freedom to replace $T_{1/1}^\nu$ by
$$T_{1/1\>B}^\nu\=d T_{1/1}^\nu +\gamma B^\nu,\quad\quad\quad\gamma\in {\bf
C}\>\>
{\rm arbitrary},\eqno(2.27)$$
if $\d_\nu B^\nu=0$. Requiring additionally $B^\nu$ to fulfil the properties
(A), (B), (C) and
(D) listed in sect.1.2, and to have ghost charge $-1$, there remains only one
possibility
$$B^\nu (x)=\d_\mu D^{\nu\mu}(x)\eqno(2.28)$$
with
$$D^{\nu\mu}(x)\=d
igf_{abc}:u_a(x)A^\nu_b(x)A^\mu_c(x):=-D^{\mu\nu}(x).\eqno(2.29)$$
This is proven in [11,13].
The $T_n$-distribution with a modified Q-vertex $T_{1/1\>B}^\nu$ (rsp. an
original Q-vertex
$T_{1/1}^\nu$/ rsp. a vertex $B^\nu$/ rsp. $D^{\nu\mu}$) at $x_l$ and with all
other vertices being
a $T_1\=d (T_1^0+T_1^2+T_1^7)$-coupling is denoted by $T_{n/l\>B}^\nu
(x_1,...,x_l,...,x_n)$
(rsp. $T_{n/l}^\nu (x_1,...,x_n)$,/ rsp. $B_{n/l}^\nu (x_1,...,x_n)$,/ rsp.
$D_{n/l}^{\nu\mu}
(x_1,...,x_n)$). The relation $D^{\nu\mu}=-D^{\mu\nu}$ can be maintained in the
inductive
construction of the $T_n$'s
$$D^{\nu\mu}_{n/l}=-D^{\mu\nu}_{n/l}.\eqno(2.30)$$
This is evident for the tensor products (A.1-3) and for the steps (A.4),
(A.13-15).
Concerning the splitting (A.7) note that the antisymmetrization
(in $\nu\leftrightarrow\mu$) of an arbitrary splitting solution
yields again a splitting solution. Due to proposition 1 (2.16), there exists a
symmetrical
normalization of $B_{n/l}^\nu$ which fulfils
$$B_{n/l}^\nu =\d^l_\mu D_{n/l}^{\nu\mu}.\eqno(2.31)$$
Moreover the normalizations can be chosen such that (2.27) propagates to higher
orders
$$T_{n/l\>B}^\nu=T_{n/l}^\nu+\gamma B_{n/l}^\nu.\eqno(2.33)$$
We conclude
$$\d_\nu^l T_{n/l\>B}^\nu =\d_\nu^l T_{n/l}^\nu.\eqno(2.33)$$
Assuming $T_n,\>T_{n/l}^\nu\>\>(l=1,...,n)$ to be gauge invariant (i.e. to
fulfil (1.8)),
there exists a symmetrical normalization of the distributions $T_{n/l\>B}^\nu$,
such
that $T_n,\>T_{n/l\>B}^\nu$ are gauge invariant, too. The modification (2.27)
of the Q-vertex
does not destroy gauge invariance in higher orders.
\vskip 0.5cm
{\it 2.4 Higher Orders with Coboundary-Coupling}
\vskip 0.5cm
The results of this subsection are summarized in

{\bf Proposition 4}: {\it Choosing suitable symmetrical normalizations the
following statements hold
true for all $F=A',R',R'',D,T,\tilde T$:

In orders $1\leq n\leq 4$ the $F_n$'s with coboundary-coupling have the
representation
$$F_n^{7...70...0}={1\over r}\{d_Q F_n^{67...70...0}+d_Q
F_n^{767...70...0}+...+d_Q
F_n^{7...760...0}\}+\eqno(2.34a)$$
$$+{i\over
r}\sum_{l=r+1}^n\d_\nu^l\{F_n^{67...70...010...0\nu}+F_n^{767...70...010...0\nu}+...
+F_n^{7...760...010...0\nu}\},\eqno(2.34b)$$
and they are gauge invariant (2.14) in orders $1\leq n\leq 3$
$$d_Q
F_n^{7...70...0}=i\sum_{l=r+1}^n\d_\nu^lF_n^{7...70...010...0\nu},\eqno(2.35)$$
where each $F_n^{...}$ has $r$ upper indices 7 or 6, $1\leq r\leq n$, and the
upper index 1 is
always at the l-th position.

These equations (2.34-35), gauge invariance (1.8) of $T_n^{0...0}$ $(n\in {\bf
N})$
and the second order identities
$$d_QF_2^{16\nu}=i\d_\mu^1F_2^{56\nu\mu}-F_2^{17\nu},\eqno(2.36)$$
$$d_QF_2^{56\nu\mu}=F_2^{57\nu\mu},\eqno(2.37)$$
$$d_QF_2^{17\nu}=i\d_\mu^1F_2^{57\nu\mu},\eqno(2.38)$$
$$d_QF_2^{10\nu}=i\d_\mu^1F_2^{50\nu\mu}-i\d_\mu^2F_2^{11\nu\mu},\eqno(2.39)$$
they all can be fulfilled simultaneously.}

Remarks: (1) Replacing $F_n^{i_1...i_n}(x_1,...,x_n)$ by
$T_1^{i_1}(x_1)...T_1^{i_n}(x_n)$
and applying (1.11), (1.13), (2.7-9), (2.11) and (2.13), the equations
(2.34-39) are
obviously fulfilled - this is the intuition.

(2) Due to (2.19), similar equations with permuted upper indices hold true for
$T_n,\,\tilde T_n$.

(3) Applying $d_Q$ to (2.34) we obtain
$$d_Q F_n^{7...70...0}={i\over
r}\sum_{l=r+1}^n\d_\nu^l\{d_QF_n^{67...70...010...0\nu}+...
+d_QF_n^{7...760...010...0\nu}\}.\eqno(2.40)$$
However, this is not gauge invariance in the sense of Q-vertices (2.14). The
latter is
given by (2.35).

(4) By means of (2.34-35) the list (2.36-39) of second order identities, which
are a kind of
gauge invariance equations, can be extended
$$d_QF_2^{70}=i\d_\nu^2F_2^{71\nu},\eqno(2.41)$$
$$d_QF_2^{77}=0,\eqno(2.42)$$
$$d_QF_2^{60}=F_2^{70}-i\d_\nu^2F_2^{61\nu}.\eqno(2.43)$$
$${1\over 2}(d_QF_2^{67}+d_QF_2^{76})=F_2^{77}.\eqno(2.44)$$

{\it Proof of proposition 4: (A) Outline}: The proof of (2.34-35) is by
induction on the order $n$.
However, we will see that the proof of (2.35) in order $n$ needs identities of
the type (2.36),
(2.38-39) in lower orders $k\leq n-1$.
But (2.39) cannot be proven by means of the general, elementary inductive
methods
of this section, it needs an explicit proof which uses the actual couplings
(1.2-4), (1.7) and (2.5).
This proof, which is given in appendix B, is similar to the proof of gauge
invariance (1.8) of
$T_2^{00}$. To prove an identity analogous to (2.39) in higher orders (see
(2.50a) below), requires
a huge amount of work (compare [2-5]), which is not done in this paper.
Therefore, the inductive proof of gauge invariance (2.35) stops at $n=3$.
Moreover the proof of (2.34) in order $n$ needs (2.35) in lower orders $k\leq
n-1$.
Consequently, the representation (2.34) of $F_n^{7...70...0}$ will be proven
for $n\leq 4$ only.

{\it (B) Proof of (2.34) by means of (2.34-35) in lower orders}: We start with
(A.2)
$$A_n^{\prime 7...70...0}(x_1,...;x_n)=\sum_{X,Y}\{{s\over r}\tilde
T_k^{7...70...0}(X)
T_{n-k}^{7...70...0}(Y,x_n)+\eqno(2.45a)$$
$$+{r-s\over r}\tilde T_k^{7...70...0}(X)
T_{n-k}^{7...70...0}(Y,x_n)\},\eqno(2.45b)$$
where $\tilde T_k^{7...70...0}$ (rsp. $T_{n-k}^{7...70...0}$) has $s$ (rsp.
$r-s$) upper
indices 7. Next we insert the induction hypothesis (2.34) for $\tilde
T_k^{7...70...0}$
into (2.45a) (rsp. (2.34) for $T_{n-k}^{7...70...0}$ into (2.45b)). Then we
apply (1.11) to the terms
with a $d_Q$-operator and obtain
$${s\over r}\tilde T_k^{7...70...0}(X) T_{n-k}^{7...70...0}(Y,x_n)={1\over
r}\Bigl [
d_Q(\tilde T_k^{67...70...0}(X) T_{n-k}^{7...70...0}(Y,x_n))+\eqno(2.46a)$$
$$+\tilde T_k^{67...70...0}(X)
d_QT_{n-k}^{7...70...0}(Y,x_n)+...+\eqno(2.46b)$$
$$+i\sum_{l=s+1}^k\{(\d_\nu^l\tilde
T_k^{67...70...010...0\nu}(X))T_{n-k}^{7...70...0}(Y,x_n)
+...\}\Bigl ],\eqno(2.46c)$$
and similar for (2.45b). The next step is to insert the induction hypothesis
(2.35)
or gauge invariance (1.8) (the latter in the special case $r-s=0$) into
$d_QT_{n-k}^{7...70...0}(Y,x_n)$ in (2.46b). Then we see that the
$A'_n$-distributions fulfil
(2.34): The terms of type (2.46a) add up to (2.34a); (2.46b) and (2.46c) can be
combined and
all terms of this type give together (2.34b). Similarly one proves that the
$R'_n$-, $R''_n$-
and, therefore, the $D_n$-distributions satisfy (2.34).

We turn to the splitting (A.7). Let
$R_n^{67...70...0},\,R_n^{767...70...0},...,
R_n^{67...70...010...0\nu},\break R_n^{767...70...010...0\nu},...$ be arbitrary
splitting solutions
of the corresponding $D_n^{...}$-distributions. By means of the definition
$$R_n^{7...70...0}\=d {1\over r}\{d_Q R_n^{67...70...0}+d_Q
R_n^{767...70...0}+...\}+$$
$$+{i\over
r}\sum_{l=r+1}^n\d_\nu^l\{R_n^{67...70...010...0\nu}+R_n^{767...70...010...0\nu}+...
\},\eqno(2.34')$$
we obtain a splitting solution of $D_n^{7...70...0}$, analogously to (2.22),
(2.26). Obviously
(2.34) is maintained in the remaining steps - the construction of $T'_n,\,T_n$
and $\tilde T_n$
(A.13-15).

{\it (C) Proof of (2.35) by means of (2.34) in the same order $n$, and by means
of (2.34-35)
and identities of the type  (2.36), (2.38-39) in lower orders}: One easily
verifies (by inserting
(2.35) and (1.8) in lower orders) that the $A'_n$-, $R'_n$-,
$R''_n$-distributions fulfil (2.35).
Therefore,
as usual gauge invariance (2.35) can be violated in the distribution splitting
only. However,
to prove that this violation can be avoided by choosing a suitable
normalization, is a
completely non-trivial business [1-5]. Moreover the normalization of
$T_n^{7...70...0}$
is restricted by $(2.34')$. Therefore, we go another way to prove (2.35) for
$T_n,\,\tilde T_n$.
We show that the r.h.s. of (2.40) agrees with the r.h.s. of (2.35), if a
suitable
symmetrical normalization of $T_n^{7...70...010...0\nu},\>1\leq r\leq n-1,$ is
chosen.
(The case $r=n$ is trivial.) For this purpose we consider
$$A_n^{\prime 7...70...010...0\nu}-{1\over r}\{d_QA_n^{\prime
67...70...010...0\nu}+...
+d_QA_n^{\prime 7...760...010...0\nu}\},\eqno(2.47)$$
where the upper index 1 is always at the $l$-th position. We insert the
definition (A.2) of
the $A'_n$-distributions. Similarly to (2.25) we
then apply (1.11) and the induction hypothesis, i.e. we insert
(2.7-8), (2.11) and (2.13) if $n=2$, and additionally (2.36), (2.38-39),
(2.41-44) if $n=3$.
In this way we obtain
$$(2.47)={i\over r}\Bigl\{\sum_{j=r+1\>(j\not= l)}^n[\pm\d_\mu^jA_n^{\prime
67...70...010...010...0\mu\nu}\pm ...
\pm\d_\mu^j A_n^{\prime 7...760...010...010...0\mu\nu}]+\eqno(2.48a)$$
$$+\d_\mu^lA_n^{\prime 67...70...050...0\nu\mu}+...+\d_\mu^lA_n^{\prime
7...760...050...0\nu\mu}
\Bigl\}.\eqno(2.48b)$$
In (2.48a) the two upper indices 1 are at the $j$-th and $l$-th position, and
we have a plus
(rsp. a minus) if $j<l$ (rsp. $j>l$). One proves (2.47)=(2.48) for the $R'_n$-,
$R''_n$-distributions
in a similar way.

Analogously to (2.30), the antisymmetry $T_1^{5\nu\mu}=-T_1^{5\mu\nu}$ (2.5)
can be preserved
in the inductive construction of the $T_n$'s. Starting with arbitrary splitting
solutions
$R_n^{67...70...050...0\nu\mu}=-R_n^{67...70...050...0\mu\nu},...,R_n^{7...760...050...0\nu\mu}
=-R_n^{7...760...050...0\mu\nu},\break R_n^{67...70...010...010...0},...,
R_n^{7...760...010...010...0}$ we may (similar to $(2.34')$) define
$R_n^{7...70...010...0}$
by the equation
(2.47)=(2.48) (with $A_n^{\prime ...}$ everywhere replaced by $R_n^{...}$).
This equation
is not destroyed in the construction of $T'_n,\,T_n$ and $\tilde T_n$. Summing
up we have proven
$$F_n^{7...70...010...0\nu}-{1\over r}\{d_QF_n^{67...70...010...0\nu}+...
+d_QF_n^{7...760...010...0\nu}\}+$$
$$={i\over r}\Bigl\{\sum_{j=r+1\>(j\not=
l)}^n[\pm\d_\mu^jF_n^{67...70...010...010...0\mu\nu}\pm ...
\pm\d_\mu^j F_n^{7...760...010...010...0\mu\nu}]+$$
$$+\d_\mu^lF_n^{67...70...050...0\nu\mu}+...+\d_\mu^lF_n^{7...760...050...0\nu\mu}\Bigl\}\eqno(2.49)$$
for all $F=A',R',R'',D,A,R,T',T,\tilde T$ and for $n\leq 3,\>1\leq r\leq n-1$.
We insert this equation into
$$\sum_{l=r+1}^n\d_\nu^l\Bigl\{F_n^{7...70...010...0\nu}-{1\over
r}[d_QF_n^{67...70...010...0\nu}+...
+d_QF_n^{7...760...010...0\nu}]\Bigl\},\eqno(2.50)$$
for $F=T,\tilde T$. Taking the different signs of the $(j,l)$- and the
$(l,j)$-term in
$\sum_{j,l\>(j\not= l)}\pm\d^l\d^j\break F_n^{...010...010...}$ and
$F_n^{...5...\nu\mu}=-F_n^{...5...\mu\nu}$ into account, we see that (2.50)
vanishes. This is
the desired result.

{\it Proof of (2.36-39)}: The first identity (2.36) is the case $n=2,\,r=1$ of
(2.49). All these
equations (2.36-39) are easily verified for the $A_2^{\prime
...}$-distributions etc.
and, therefore, can be violated in the splitting only. The latter is no problem
for
(2.37), since we may define $R_2^{57\nu\mu}\=d d_QR_2^{56\nu\mu}$ for an
arbitrary splitting
solution $R_2^{56}$. Applying $d_Q$ to (2.36), we obtain (2.38) by means of
(2.37).
It remains (2.39), which is proven in appendix (B) by explicitly inserting the
actual
couplings. It turns out that
there exists a normalization of $T_2^{10\nu}(x_1,x_2)=T_2^{01\nu}(x_2,x_1)$
such that (2.39)
and gauge invariance (1.8) (in second order) are satisfied simultaneously. One
easily verifies
that this is the only problem of compatibility in (2.34-39) and (1.8). For
example in second order
the distributions
$T_2^{56\nu\mu}=-T_2^{56\mu\nu},\>T_2^{61},\>T_2^{60},\>T_2^{67}$ can
be normalized in an arbitrary symmetrical way. Then the normalizations of
$T_2^{17},\>T_2^{57},
\>T_2^{70},\>T_2^{77}$ are uniquely fixed by (2.36-37), (2.43-44), and all
identities
(2.36-38) and (2.41-44) are fulfilled. The remaining distributions
$T_2^{00},\>T_2^{10},
\>T_2^{50}$ and $T_2^{11}$ appear in (1.8) and (2.39) only. $\quad\quad\w$

If the identities ($F=T,\,\tilde T$)
$$d_QF_n^{5...51...10...0}=i\sum_{j=t+1}^{t+s}(-1)^{(j-t-1)}\d^jF_n^{5...51...151...10...0}+$$
$$+i(-1)^s\sum_{j=t+s+1}^n\d^jF_n^{5...51...10...010...0},\quad\quad n\in {\bf
N},
\>\quad 0\leq t,s\leq n,\>\quad t+s\leq n\eqno(2.50a)$$
hold true (where $F_n^{5...51...10...0}$ on the l.h.s. has $t$ indices 5, $s$
indices 1 and all
derivatives on the r.h.s. are divergences, the Lorentz indices are omitted),
one can prove the representation (2.34) and gauge
invariance (2.35) in all orders. This is shown in appendix C by a
generalization of this
proof here. Unfortunately an inductive proof of (2.50a) by means of the simple
technique of
this section fails because of the splitting (A.7) - there is no term in (2.50a)
which has neither
a $d_Q$-operator nor a derivative. We emphasize that the identities (2.50a) do
not depend on the
explicit form (1.14-15) of the coboundary coupling (no upper indices 6 or 7
appear in (2.50a)).
These identities concern solely the starting-coupling $T_1^0$, its Q-vertex
$T_1^1$
and its Q-Q-vertex $T_1^5$.

Remark: The compatibility of (2.39) and gauge invariance (1.8) in second order
is remarkable
in the tree sector: Each of this two identities fixes the normalization of
$T_2^{10}\vert_{\rm tree}$
{\it uniquely} and these two normalizations agree in fact (see appendix B and
sect.3.2 of [5]).
This is a further hint that our gauge invariance (1.8) relies on a deeper
(cohomological ?)
structure. The knowledge of the latter would presumably shorten the proof of
(1.8) and would
be an excellent tool to prove the missing identities (2.50a).
\vskip 0.5cm
{\it 2.5 Generality of the Results}
\vskip 0.5cm
In the preceeding subsections 2.2 and 2.4 the explicit structures of the
starting-theory $T_1^0$
(1.2), of the corresponding Q-vertex $T_1^{1\nu}$ (1.7), of the
divergence-coupling (1.16-17)
and of the coboundary-coupling (1.14-15) have not been needed. We have solely
used the
following properties:

(i) The starting-theory $T_1^0$ is gauge invariant with respect to the Q-vertex
$T_1^{1\nu}$
in all orders which are considered.

(ii) There exists a Q-Q-vertex $T_1^{5\nu\mu}(x)$ which fulfils
$$T_1^{5\nu\mu}=-T_1^{5\mu\nu},\quad\quad d_QT_1^{5\nu\mu}=0\quad\quad\quad
{\rm and}\quad\quad
d_QT_1^{1\nu}(x)=i\d_\mu T_1^{5\nu\mu}(x).\eqno(2.51)$$

(iii) The second order identity (2.39) holds true and is compatible with gauge
invariance (1.8)
of $T_2^{00}$.

Only (i) is needed in sect.2.2. Therefore, the results about the
divergence-coupling
apply to any gauge invariant quantum field theory, e.g. to quantum gravity
[14]. This holds also true
for (2.34) in second order, i.e. (2.43-44).

If additionally (ii) is fulfilled ($d_QT_1^5=0$ is not needed for the following
statement),
gauge invariance (2.35) is proven in second order (i.e. (2.41-42) are valid),
and this implies
the identities (2.34) up to third order.
Note that the modified Q-vertex $T_{1/1\>B}^\nu$ (2.27) satisfies (ii), too,
$$d_QT_{1/1\>B}^\nu =i\d_\mu T_{1\>B}^{5\nu\mu}\eqno(2.52)$$
with
$$T_{1\>B}^{5\nu\mu}\=d T_1^{5\nu\mu} -i\gamma d_Q
D^{\nu\mu}=-T_{1\>B}^{5\mu\nu},
\quad\quad d_QT_{1\>B}^{5\nu\mu}=0.\eqno(2.53)$$

For a model which satisfies (i), (ii) and all identities (2.50a) ((2.39) is a
special case of
the latter) also the statements (2.34-35) about the coboundary-coupling are
proven in all orders.
\vskip 0.5cm
{\it 2.6 $n$-Point Distributions with Divergence- and Coboundary-Coupling}
\vskip 0.5cm
The general case (2.10) of $T_n$ containing the ordinary Yang-Mills coupling
$T_1^0$, the divergence-
{\it and} the coboundary-coupling can easily
be traced back to the results of the preceeding sections 2.2 and 2.4-5. We
replace $T_1^0$ by
$$\bar T_1^0\=d T_1^0+T_1^2=T_1^0+\beta_2 \d_\nu K_2^\nu \eqno(2.54)$$
and $T_1^{1\nu}$ by
$$\bar T_1^{1\nu}\=d T_1^{1\nu}+T_1^{3\nu}=T_1^{1\nu}-i\beta_2
d_QK_2^\nu.\eqno(2.55)$$
Due to corollary 3, the $\bar T_1^0$-theory is gauge invariant with respect to
the Q-vertex
$\bar T_1^{1\nu}$ in all orders, i.e. property (i) of subsect.2.5 is fulfilled.
Obviously
property (ii) holds also true with the old $T_1^5$-vertex (2.5): $d_Q\bar
T_1^{1\nu}=i\d_\mu
T_1^{5\nu\mu}$. It would be very suprising if (2.39) would be wrong for the
$(\bar T_1^0,\>
\bar T_1^{1\nu},\>T_1^{5\nu\mu})$-couplings. By means of proposition 4 we
conclude that the general
$n$-point distributions (2.10) (with coboundary- {\it and} divergence-coupling)
are gauge invariant
in second and most probably third order, and we obtain the representation
(2.34) with
respect to the coboundary-vertices up to third (rsp. fourth) order.

Let us describe an alternative way. We replace $T_1^0$ by
$$\bar T_1^0\=d T_1^0+T_1^7=T_1^0+\beta_1 d_QK_1.\eqno(2.56)$$
The Q-vertex (1.7) needs no change: $d_Q \bar T_1^0=i\d_\nu T_1^{1\nu}$.
Proposition 4
(2.35) tells us that the $\bar T_1^0$-theory is gauge invariant up to third
order.
Applying corollary 3 we obtain gauge invariance (2.14) of the general $T_n$'s
(2.10) up to third order.
Moreover, due to proposition 1, these distributions are divergences with
respect to their
divergence-vertices in any order.
\vskip 0.5cm
{\it 2.7 Gauge Invariant Normalization of Second Order Tree Diagrams}
\vskip 0.5cm
We only consider the tree sector and start with the following normalization of
$T_2(x_1,x_2)$ (2.10)
($T_2\=d
T_2^{00}+T_2^{20}+T_2^{02}+T_2^{22}+T_2^{70}+T_2^{07}+T_2^{77}+T_2^{27}+T_2^{72}$):
The C-number distributions of $T_{20}\vert_{\rm tree}$ (the lower index 0
indicates this special
normalization) are
$$t_{\cal O}(x_1-x_2)\sim D^F(x_1-x_2),\>\d^\mu D^F(x_1-x_2),\>\d^\mu\d^\nu
D^F(x_1-x_2),\eqno(2.57)$$
they have no local terms.
The singular order $\omega$ of $t_{\cal O}$ (rsp. the number of derivatives on
$D^F$ in (2.57)) can be
computed from the combination ${\cal O}$ of the four external free field
operators (see
$\omega ({\cal O})$ in (A.17)) and is $\omega ({\cal O}) =-2,-1,0$. For each
four-legs
combination ${\cal O}$ with $\omega ({\cal O}) =0$ we may add a local term
$$N_{\cal O}(x_1-x_2)=C_{\cal O}\delta (x_1-x_2):{\cal
O}(x_1-x_2):\eqno(2.58)$$
to $T_{20}$, where $C_{\cal O}$ is a free normalization constant (A.12).
Gauge invariance (2.14) fixes the values of $C_{\cal O}$
uniquely [1,5,13]. In $T_2^{00}$ the normalization term
$$N_{AAAA}(x_1-x_2)=-{1\over 2}ig^2f_{abr}f_{cdr}\delta (x_1-x_2)
:A_{\mu a}A_{\nu b}A_c^\mu A_d^\nu:\eqno(2.59)$$
is required [1,5]. This is the four-gluon interaction, which propagates to
higher orders
in the inductive construction of the $T_n$'s (sect.4.2 of [15]). The
normalization terms (2.58)
of $T_2^{20},...,T_2^{72}$ which are needed for gauge invariance (2.14)
can quickly be calculated by using our results. We have proven that
$T_2^{20},\>T_2^{22},\>
T_2^{70},\>T_2^{77}$ and $T_2^{27}$ are gauge invariant with the normalizations
given by
proposition 1 (2.16), rsp. proposition 4 (2.34). (In the case of $T_2^{27}$ we
do the replacement
(2.54-55) (or alternatively (2.56)) before applying (2.34) (rsp. (2.16))).
Therefore, we simply
have to pick out the local terms in
$\d_\mu^1T_2^{40\mu}(=T_2^{20}),\>\d_\mu^1\d_\nu^2T_2^{44\mu\nu}
(=T_2^{22}),\>d_QT_2^{60}+i\d_\nu^2T_2^{61\nu}(=T_2^{70})$ and in ${1\over
2}(d_QT_2^{67}+d_QT_2^{76})
(=T_2^{77})$. In the tree sector there are no local terms in
$T_2^{40},\>T_2^{44},\>T_2^{60},\>
T_2^{61},\>T_2^{67}$ (their normalization is unique)
and, therefore, neither in $d_QT_2^{60},\>d_QT_2^{67}$. All local terms
are generated by the divergences in
$\d_\mu^1T_2^{40\mu},\>\d_\mu^1\d_\nu^2T_2^{44\mu\nu}$ or
$i\d_\nu^2T_2^{61\nu}$, due to $\w D^F(x_1-x_2)=\delta (x_1-x_2)$. It turns out
that all these
local terms are four-ghost interactions, which add up to
$$N_{uu\tilde u\tilde u}(x_1-x_2)=-ig^2\bigl ({(\beta_2)^2\over
2}+\beta_1-2\beta_1\beta_2\bigl )
f_{abr}f_{cdr}\delta (x_1-x_2):u_au_b\tilde u_c\tilde u_d:\eqno(2.60)$$
in agreement with the much longer calculation in [13].

Remarks: (1) The powers of $\beta_1,\,\beta_2$ in (2.60) tell us the origin of
the corresponding term.
For example the term $\sim \beta_1\beta_2$ comes from $T_2^{27}+T_2^{72}$.

(2) We have seen that on the tree sector the normalizations of
$T_2^{20},...,T_2^{72}$ are {\it
uniquely} fixed by (2.16) or (2.34). However, this does not imply that gauge
invariance fixes
the normalization of $T_2^{20}\vert_{\rm tree},...,T_2^{72}\vert_{\rm tree}$
uniquely.
The latter statement is a by-product of the calculation in [13].

(3) In agreement with our observations in first order (see remark (1) in
sect.1.2),
there is {\it no ambiguity in the four-gluon interaction (2.59) -
it is independent on $\beta_1,\,\beta_2$.}

(4) The most general coupling which is gauge invariant (2.14) in all orders
(this is not proven
completely for the coboundary-coupling) has been given. It can be compared with
the most
general Lagrangian (written in terms of interacting fields) which is invariant
under the
full BRS-transformations of the interacting fields - see formula (3.13) of
[10].
For this purpose we must choose the Feynman gauge $\lambda =1$ in this formula.
Then one easily
verifies that the terms $\sim g$ and $\sim g^2$ in the interaction part of this
Lagrangian
agree with $(T_1^0+\beta_1d_QK_1+\beta_2\d_\mu K_2^\mu)\sim g$ and with
$N_{AAAA},\,
N_{uu\tilde u\tilde u}\sim g^2$, if we set $\beta_2=2\beta_1$ and identify the
free parameter $\alpha$
of [10] with $\beta_2=2\beta_1$. There is only a one-parametric freedom in [10]
which is
given by adding to the Lagrangian $\alpha s(...)$. The latter is a coboundary
with respect to the
BRS-operator $s$. In doing so the Lagrangian remains $s$-invariant, due to the
nilpotency of $s$.
This seems to be analogous to our coboundary-coupling $\beta_1d_QK_1$ (1.14).
However, we see
from $\alpha=2\beta_1=\beta_2$ that there is not a complete correspondence - a
change of $\alpha$
means also the addition of a divergence $\beta_2\d_\mu K_2^\mu$ (1.16). Since
in our framework
the interaction is switched off by $g\in {\cal S}({\bf R}^4)$, our gauge
invariance is not
$[Q,T_n]=0$ but $[Q,T_n]=$(divergences), and, therefore, we have the freedom of
adding a
divergence-coupling (1.16) to $T_1$. This explains the fact that we have a
two-parametric
freedom and not only a one-parametric one.

(5) We call a normalization term $N_{\cal O}$ (2.58) 'natural', if there is a
corresponding
non-vanishing non-local term, more precisely if $T_{20}\vert_{\rm tree}$ (2.57)
contains a non-vanishing C-number distribution $t_{\cal O}$ (with the same
${\cal O}$).
$N_{AAAA}$ (2.59) is of this kind. It can be generated by replacing
$$\d^\mu\d^\nu D^F(x_1-x_2)\quad\quad\quad{\rm by}\quad\quad\quad [\d^\mu\d^\nu
D^F(x_1-x_2)-
{1\over 2}g^{\mu\nu}\delta (x_1-x_2)]\eqno(2.61)$$
in $t_{AAAA}$ [1,5]. The other normalization terms are called 'unnatural',
since they do not
naturally arise in the inductive construction of the $T_n$'s - the numerical
distribution
$d_{\cal O}=0$ is splitted in $d_{\cal
O}(x_1-x_2)=\delta^{(4)}(x_1-x_2)-\delta^{(4)}(x_1-x_2)$.
$N_{uu\tilde u\tilde u}$ is unnatural,
because in the corresponding diagram $\d_\mu A^\mu_a(x_1)$ and $\d_\nu
A^\nu_b(x_2)$ are contracted,
which gives $-i\delta_{ab}g^{\mu\nu}\d_\mu\d_\nu D^+_0(x_1-x_2)=0$.
($D^+_0(x_1-x_2)$ is the positive
frequency part of the massless Pauli-Jordan distribution.) Note that the proof
of gauge invariance
(1.8) in higher orders $n\geq 3$ [2-5] uses normalizations which could be
unnatural in an
analogous sense.
\vskip 0.5cm
{\it 2.8 Non-Uniqueness of Quantized Yang-Mills Theories}
\vskip 0.5cm
To simplify the discussion we assume (2.34) and (2.35) to hold true in any
order. Then the ambiguities
of quantized Yang-Mills theories, which are given by the free choice of the
parameters
$\beta_1,\beta_2\in{\bf R}$ (1.14), (1.16), are not restricted by gauge
invariance in higher
orders, due to corollary 3 and (2.35). The freedom is reduced to a
one-parametric set,
if we admit only natural normalization terms for second order tree diagrams
$$N_{uu\tilde u\tilde
u}=0\quad\Longleftrightarrow\quad\beta_1={(\beta_2)^2\over 4\beta_2-2},
\>\>\>\beta_2\not= {1\over 2}.\eqno(2.62)$$
This prescription agrees partially with the Faddeev-Popov procedure:
The exponentiation of a determinant can generate only terms quadratic in the
ghosts.
Therefore, the Faddeev-Popov method cannot yield a four-ghost interaction.

There is a more technical criterion which gives another restriction of the
ambiguities and
roughly speaking requires that the cancellations in the gauge invariance
equation (2.14)
are {\it simple}. To be more precise let us consider this equation
for second order tree diagrams. In the natural operator
decomposition [5] the terms $\sim\d^\mu\delta (x_1-x_2)$ cancel completely iff
$$\beta_2=0.\eqno(2.63)$$
(For $\beta_2\not= 0$ the terms $\sim\d\delta :{\cal O}:$ must be combined with
terms
$\sim\delta :{\cal O'}:$, where the difference of the two operator combinations
${\cal O'}$ and ${\cal O}$ is that ${\cal O'}$ has one derivative more.)
Let us assume that one can prove C-number identities (called 'Cg-identities'
[2-5]) which express
gauge invariance (2.14). Then the transition from the natural operator
decomposition of (2.14)
to the Cg-operator decomposition (i.e. the op. dec. in which the Cg-identities
hold true)
is much more complicated for $\beta_2\not= 0$ than for $\beta_1=0=\beta_2$ [5].
We see
from (2.62-63) that {\it the theory with $\beta_1=0=\beta_2$ is the most simple
one}. However,
this does not exclude the other values of $\beta_1,\beta_2$, since we can
construct a Lorentz-,
SU(N)- and P-, T-, C-invariant, (re)normalizable, gauge invariant and
pseudo-unitary S-matrix
for any choice of $\beta_1,\beta_2\in {\bf R}$.

We turn to the physical consequences of the freedom in the choice of
$\beta_1,\beta_2$.
For this purpose we consider $PT_n(x_1,...,x_n)P$, where $T_n$ is given by
(2.10) and
$P$ is the projector on the physical subspace [4]. By means of
$d_QA_a^\mu=i\d^\mu
u_a,\>d_Qu_a=0,\>d_Q\tilde u_a=-i\d_\nu A^\nu_a$ and the fact that $\d^\mu u_a$
and
$\d_\nu A^\nu_a$ are unphysical fields, we conclude
$$Pd_QF_n(x_1,...,x_n)P=0,\eqno(2.64)$$
where $F=A',R',R'',D,A,R,T',T,\tilde T$. Together with propositions 1 and 4
(2.16),
(2.34) we obtain
$$PT_n(x_1,...,x_n)P=T_n^{0...0}(x_1,...,x_n)+({\rm
sum\>\>of\>\>divergences}).\eqno(2.65)$$
On the r.h.s. {\it the dependence on $\beta_1,\beta_2$ is exclusively in the
divergences}. But
the infrared behavior of Yang-Mills theories is not under control. Therefore,
we cannot conclude
that the divergences in (2.65) vanish in the adiabatic limit $g\rightarrow 1$.
\vskip 1cm
{\trm Appendix A: Inductive Construction of the $T_n$'s\break\vskip 0.2cm
according to Epstein and Glaser}
\vskip 1cm
The input of the inductive construction of the $T_n$'s (1.1) are the $T_1^i$'s
(e.g. (1.2-4), (1.7),
(2.2-7)) in terms of {\it free fields}. The couplings $T_1^i$ are roughly
speaking given by the
interaction Lagrangian densities. Let us summarize the inductive step as a
recipe. For the derivation
of this construction from causality and translation invariance
(only these two requirements are needed)
we refer the reader to [6,7]. In analogy to (1.1) we denote the n-point
distributions of the inverse
S-matrix $S(g_0,...,g_l)^{-1}$ by $\tilde T_n(x_1,...,x_n)$.
Having constructed all $T_k$, $\tilde T_k$
in lower orders $k\leq n-1$, we can define the
operator-valued distributions $R'_n,\,A'_n,\,R''_n$, which are sums of tensor
products,
$$R'_n(x_1,...;x_n)\=d\sum_{X,Y}T_{n-k}(Y,x_n)\tilde T_k (X),\eqno (A.1)$$
$$A'_n(x_1,...;x_n)\=d\sum_{X,Y} \tilde T_k (X) T_{n-k}(Y,x_n),\eqno (A.2)$$
$$R''_n(x_1,...;x_n)\=d\sum_{X,Y} T_k (X)\tilde T_{n-k}(Y,x_n),\eqno (A.3)$$
where $X\=d\{x_{i_1},...,x_{i_k}\},\quad Y\=d\{x_{i_{k+1}},...
,x_{i_{n-1}}\},\quad X\cup Y =\{x_1,...,x_{n-1}\}$ and the sum
is over all partitions of this kind with $1\leq k\equiv
\mid X\mid \leq n-1$. In order to simplify the notations, the Lorentz indices
and the upper indices
$i_s$ denoting the kind of vertex $T_1^{i_s}(x_s)$ (see e.g. (2.1-7)) are
omitted. This gives
no confusion since $i_s$ is strictly coupled to the time-space argument $x_s$.
One can prove that
$$D_n\=d R'_n-A'_n\eqno(A.4)$$
has causal support
$${\rm supp}\,D_n(x_1,...;x_n)\subset
(\Gamma^+_{n-1}(x_n)\cup\Gamma^-_{n-1}(x_n)),\eqno(A.5)$$
where
$$\Gamma^\pm_{n-1}(x_n)\=d\{(x_1,...,x_n)\in {\bf R}^{4n}|
x_j\in x_n+\bar V^\pm,\>\forall j=1,...,n-1\}.\eqno(A.6)$$
The crucial step in the inductive construction is the {\it correct
distribution splitting} of $D_n$
$$D_n=R_n-A_n,\eqno(A.7)$$
with
$${\rm supp}\,R_n(x_1,...;x_n)\subset \Gamma^+_{n-1}(x_n)\quad
{\rm and}\quad
{\rm supp}\,A_n(x_1,...;x_n)\subset \Gamma^-_{n-1}(x_n).\eqno(A.8)$$
For this purpose we expand the operator-valued distributions
in normally ordered form
$$F_n(x_1,...,x_n)=\sum_{{\cal O}}f_{\cal O}(x_1-x_n,...,x_{n-1}-x_n)
:{\cal O}(x_1,...,x_n):,\eqno(A.9)$$
where $F=R',\,A',\,D,\,R,\,A,\,T,\tilde T$ and ${\cal O}(x_1,...,x_n)$ is a
combination of the free field operators. The coefficients $f_{\cal O}$
are C-number distributions. Due to translation invariance,
they depend on the relative coordinates only and, therefore, are
responsible for the support properties. Consequently,
the splitting must be done in these C-number distributions.
Obviously, the critical point for the splitting is the UV-point
$$\Gamma^+_{n-1}(x_n)\cap \Gamma^-_{n-1}(x_n)=
\{(x_1,...,x_n)\in {\bf R}^{4n}|x_1=x_2=...=x_n\}.\eqno(A.10)$$
In order to measure the behavior of the C-number distribution
$f$ in the vicinity of this point, one defines an index $\omega (f)$,
which is called the {\it singular order of $f$ at $x=0$} [6,7].
We will need the following example: Let $D^a,\>a\=d (a_1,...,a_m)$,
be a partial differential operator. Then
$$\omega (D^a\delta^{(m)}(x_1,...,x_m))=|a|\=d a_1+...+a_m.\eqno(A.11)$$

If $\omega (d_{\cal O})<0$, the splitting of $d_{\cal O}$ is trivial and
uniquely given
by multiplication with a step function [6,7].

If $\omega (d_{\cal O})\geq 0$, one must do the splitting more carefully
[6,7]. Moreover it is not unique. One has an undetermined
polynomial which is of degree $\omega(d_{\cal O})$ (the degree cannot be higher
since renormalizability requires $\omega(r_{\cal O})= \omega(d_{\cal O})$),
$$r_{\cal O}(x_1-x_n,...,x_{n-1}-x_n)=r_{\cal O}^0(...)
+\sum_{|a|=0}^{\omega (d_{\cal O})}C_aD^a\delta^{(4(n-1))}
(x_1-x_n,...,x_{n-1}-x_n),\eqno(A.12)$$
where $r_{\cal O}^0$ is a special splitting solution and
$C_a$ are the undetermined normalization constants.
If one does the splitting also in this case by multiplying with a
step function, one obtains the usual, UV-divergent Feynman
rules. But this procedure is mathematically inconsistent.
The correct distribution splitting saves us from UV-divergences.

From $R_n$ one constructs
$$T'_n\=d R_n-R'_n\eqno(A.13)$$
and $T_n$ is obtained by symmetrization of $T'_n$
$$T_n^{i_1...i_n}(x_1,...,x_n)=\sum_{\pi\in {\cal S}_n}{1 \over n!}
T_n^{\prime i_{\pi 1}...i_{\pi n}} (x_{\pi 1},...,x_{\pi n}).\eqno(A.14)$$
In order to finish the inductive step we must construct
$$\tilde T_n\=d -T_n-R'_n-R''_n.\eqno(A.15)$$
One can prove that (A.14-15) are the correct n-point distributions of
$S(g_0,...,g_l)$ (1.1) rsp. $S(g_0,...,g_l)^{-1}$, fulfilling the requirements
of causality and
translation invariance. Note
$$\omega\=d\omega (t_{\cal O})=\omega (r_{\cal O})=\omega (d_{\cal
O}).\eqno(A.16)$$
The undetermined local terms (A.12) go over from $r_{\cal O}$ to $t_{\cal O}$.
The normalization constants $C_a$ are restricted by Lorentz- and
SU(N)-invariance,
the permutation symmetry (2.19), discrete symmetries, pseudo-unitarity and
gauge invariance
(compare with sect.1.2). The latter restriction plays an important role in this
paper.

In our Yang-Mills model one can prove by means of scaling properties [7]
$$\omega\leq\omega ({\cal O})\=d 4-b-g-d,\eqno(A.17)$$
where $b$ is the number of gauge bosons ($A,F$), $g$ the number of ghosts
($u,\tilde u$)
and $d$ the number of derivatives ($F,\d\tilde u,...$) in ${\cal O}$. The proof
of (A.17) in [2]
is written for $T_n^{0...0}$ and $T_n^{10...0}$. However, it goes through
without change for all
$T_n^{i_1...i_n}$ with $i_1,...,i_n\in\{0,1,2,3,5,7\}$ (see (2.1-7) for the
notations), especially
for the physically relevant $T_n$ (2.10). The couplings $T_1^4$ and $T_1^6$
have mass-dimension 3 instead of 4. Therefore, there exists a lower upper bound
$\tilde\omega
({\cal O})$ for the singular order $\omega$ of diagrams with at least one
vertex $T_1^4$ or $T_1^6$:
$\omega\leq\tilde\omega ({\cal O})<\omega ({\cal O})=4-b-g-d$. The fact that
$\omega$ is
bounded in the order $n$ of the perturbation series (here it is even
independent on $n$),
is the {\it (re)normalizability} of the model.
\vskip 1cm
{\trm Appendix B: Proof of (2.39)}
\vskip 1cm
Since (2.39) is a gauge invariance equation, it can be violated only in the
splitting (A.7) and
solely by local terms. No vacuum diagrams appear in (2.39).
\vskip 0.5cm
{\it B.1 Tree Diagrams}
\vskip 0.5cm
We work with the technique of [1]. The splitting $D_2^{...}\vert_{\rm
tree}\rightarrow
R_{20}^{...}\vert_{\rm tree}$ is done by replacing everywhere $D_0(x_1-x_2)$
(which is the mass
zero Pauli-Jordan distribution) by its retarded part $D_0^{ret}(x_1-x_2)$. As
in (2.57) the
lower index $0$ in $R_{20}^{...}$ and in $T_{20}^{...}\=d
R_{20}^{...}-R_{2}^{\prime...}$ (A.13-14)
indicates this special normalization in the tree sector. Note
$\w D_0^{ret}=\delta^{(4)}$, in contrast to $\w D_0=0$. This is the reason for
the appearance
of local terms $A^\nu$ which destroy (2.39)
$$d_QR_{20}^{10\nu}\vert_{\rm tree}=i\d_\mu^1R_{20}^{50\nu\mu}\vert_{\rm tree}-
i\d_\mu^2R_{20}^{11\nu\mu}\vert_{\rm tree}-A^\nu.\eqno(B.1)$$
Picking out all local terms - they all are generated in the divergences on the
r.h.s. due to
$\w D_0^{ret}=\delta^{(4)}$ - one finds
$$A^\nu (x_1,x_2)=-g^2f_{abr}f_{cdr}\Bigl \{{1\over 2}\delta
(x_1-x_2):u_au_bA_{\mu c}F_d^{\mu\nu}:+$$
$$+{1\over 2}\d^\mu\delta (x_1-x_2):u_a(x_1)u_b(x_1)A_{\mu
c}(x_2)A_d^{\nu}(x_2):+$$
$$+[g^{\tau\mu}\d^\nu\delta (x_1-x_2)-g^{\nu\mu}\d^\tau\delta (x_1-x_2)]
:A_{\tau a}(x_1)u_b(x_1)A_{\mu c}(x_2)u_d(x_2):\Bigl \}=$$
$$=i\d_\mu^1B^{\nu\mu}(x_1,x_2)+i\d_\mu^2B^{\nu\mu}(x_1,x_2)+d_QN^\nu
(x_1,x_2),\eqno(B.2)$$
with
$$B^{\nu\mu}(x_1,x_2)\=d i{g^2\over 2}f_{abr}f_{cdr}\delta
(x_1-x_2):u_au_bA_c^\mu A_d^\nu:,
\eqno(B.3)$$
$$N^\nu (x_1,x_2)\=d -ig^2 f_{abr}f_{cdr}\delta (x_1-x_2):A_{\mu a}u_bA_c^\mu
A_d^\nu:.\eqno(B.4)$$
(Note that a term $\sim f_{abr}f_{cdr}\delta (x_1-x_2):u_au_bu_c\d^\nu\tilde
u_d:$
vanishes, due to the antisymmetry of the operator part in $a,b,c$ and the
Jacobi identity for
the $f_{...}$'s.) Obviously the symmetries
$T_{20}^{50\nu\mu}(x_1,x_2)=-T_{20}^{50\mu\nu}(x_1,x_2)$
and $T_{20}^{11\nu\mu}(x_1,x_2)=-T_{20}^{11\mu\nu}(x_2,x_1)$ are preserved in
the
finite renormalizations
$$T_{2}^{50\nu\mu}\=d T_{20}^{50\nu\mu}-B^{\nu\mu},\eqno(B.5)$$
$$T_{2}^{11\nu\mu}\=d T_{20}^{11\nu\mu}+B^{\nu\mu}.\eqno(B.6)$$
and
$$T_{2}^{10\nu}\=d T_{20}^{10\nu}+N^{\nu}.\eqno(B.7)$$
Due to (B.1-2), these $T_2^{...}$-distributions (B.5-7) satisfy (2.39) on tree
level, and one
easily verifies that (2.39) fixes the normalization of $T_2^{10}\vert_{\rm
tree}$ {\it uniquely}.

On the other hand the normalization in the tree sector of
$T_2^{10\nu}(x_1,x_2)=T_2^{01\nu}(x_2,x_1)$ is uniquely
determined by gauge invariance (1.8) in second order (see sect.3.2 of [5])
$$d_QT_2^{00}=i\d_\nu^1T_2^{10\nu}+i\d_\nu^2T_2^{01\nu},\eqno(B.8)$$
where $T_2^{00}\vert_{\rm tree}$ is
normalized by (2.59) (four-gluon interaction). These two normalizations of
$T_2^{10}\vert_{\rm tree}$
((B.7) and (B.8)) agree exactly.
\vskip 0.5cm
{\it B.2 Two-Legs Diagrams}
\vskip 0.5cm
We denote the numerical two-legs distributions in the following way
$$F_{2}^{10\nu}(x_1,x_2)\vert_{\rm
2-legs}=f^{10\nu\mu}_{uA}(x_1-x_2):u_a(x_1)A_{\mu a}(x_2):+$$
$$+f^{10\nu\mu}_{Au}(x_1-x_2):A_{\mu
a}(x_1)u_a(x_2):+...:uF:+...:Fu:,\eqno(B.9)$$
$$F_{2}^{50\nu\mu}(x_1,x_2)\vert_{\rm
2-legs}=f^{50\nu\mu}_{uu}(x_1-x_2):u_a(x_1)u_a(x_2):,\eqno(B.10)$$
$$F_{2}^{11\nu\mu}(x_1,x_2)\vert_{\rm
2-legs}=f^{11\nu\mu}_{uu}(x_1-x_2):u_a(x_1)u_a(x_2):,\eqno(B.11)$$
for $(F,f)=(T,t),\,(D,d),...$ . Again we choose a normalization of
$T_{2}^{50\nu\mu}
\vert_{\rm 2-legs}$ which is antisymmetrical in $\nu\leftrightarrow\mu$.
Together with the fact
that there exists no Lorentz covariant, antisymmetric tensor of second rank
which depends on one
Lorentz vector only, we conclude
$$t^{50\nu\mu}_{uu}=0.\eqno(B.12)$$
Since $T_2^{10\nu}$ appears also in (B.8), we have some information about
$t^{10\nu\mu}_{uA},\,
t^{10\nu\mu}_{Au}$ (B.9) from the C-number identities expressing (B.8) [2],
namely
$$t^{10\nu\mu}_{Au}=-t^{10\mu\nu}_{Au}\quad\quad {\rm and\quad
therefore}\quad\quad
t^{10\nu\mu}_{Au}=0,\eqno(B.13)$$
$$\d_\nu^1 t^{10\nu\mu}_{uA}=0,\eqno(B.14)$$
$$t^{10\nu\mu}_{uA}=t^{00\nu\mu}_{AA}\quad\quad {\rm and\quad
therefore}\quad\quad
t^{10\nu\mu}_{uA}(y)=t^{10\mu\nu}_{uA}(-y),\eqno(B.15)$$
where $t^{00\nu\mu}_{AA}(x_1-x_2)$ is the C-number distribution which belongs
to the operators
$:A_{\nu a}(x_1)\break A_{\mu a}(x_2):$ in $T_2^{00}(x_1,x_2)$.
Note that $d^{11\nu\mu}_{uu}$ has exactly the same (amputated) diagrams as
$d^{10\nu\mu}_{uA}$,
consequently $d^{11\nu\mu}_{uu}=d^{10\nu\mu}_{uA}$. If we split
$d^{11\nu\mu}_{uu}$ in the same way
as $d^{10\nu\mu}_{uA}$, we obtain
$$t^{11\nu\mu}_{uu}=t^{10\nu\mu}_{uA}.\eqno(B.16)$$
Obviously (B.12-16) hold true for $t$ replaced by $\tilde t$, too.
Inserting (B.9-16) into (2.39) we see that (2.39) is fulfilled also on the
two-legs sector.
$\quad\quad\w$
\vskip 1cm
{\trm Appendix C: Coboundary-Coupling in arbitrary\break\vskip 0.2cm Order}
\vskip 1cm
To shorten the notations we shall omit the Lorentz indices and define
$${\cal S}_rF_n^{67...7i_{r+1}...i_n}\=d {1\over r}[F_n^{67...7i_{r+1}...i_n}+
F_n^{767...7i_{r+1}...i_n}+...+F_n^{7...76i_{r+1}...i_n}],\eqno(C.1)$$
where $F=T,\,\tilde T$.

{\bf Proposition 5}: {\it Assuming the identities (2.50a) to hold true, the
following equations
are simultaneously fulfilled in all orders $n\in {\bf N}$ for $F=T,\,\tilde T$,
if suitable symmetrical
normalizations are chosen:
$$d_QF_n^{7...75...51...10...0}=i\sum_{j=r+t+1}^{r+t+s}(-1)^{(j-r-t-1)}
\d^jF_n^{7...75...51...151...10...0}+$$
$$+i(-1)^s\sum_{j=r+t+s+1}^n\d^jF_n^{7...75...51...10...010...0},\quad\quad
0\leq r,t,s\leq n,\>\quad r+t+s\leq n\eqno(C.2)$$
and
$$d_Q{\cal
S}_rF_n^{67...75...51...10...0}=i\sum_{j=r+t+1}^{r+t+s}(-1)^{(j-r-t)}
\d^j{\cal S}_rF_n^{67...75...51...151...10...0}+$$
$$+i(-1)^{s+1}\sum_{j=r+t+s+1}^n\d^j{\cal
S}_rF_n^{67...75...51...10...010...0}+
F_n^{7...75...51...10...0},$$
$$1\leq r\leq n,\quad\quad 0\leq t,s\leq n,\quad\quad r+t+s\leq n,\eqno(C.3)$$
where $F_n^{7...75...51...10...0}$ and $F_n^{67...75...51...10...0}$ on the
l.h.sides have $t$
indices 5, $s$ indices 1 and $r$ indices 7, rsp. $(r-1)$ indices 7 and one
index 6. All
derivatives on the r.h.sides are divergences.}

Note that (C.2) is a generalization of gauge invariance (2.35) and (1.8); the
representation
(2.34) and (2.49) are special cases of (C.3). The indices may be permuted in
(C.2-3) according
to (2.19).

{\it Proof:} The reasoning runs essentially along the same lines as the proof
of prop.4.
Therefore, we only sketch it. First we consider (C.3). We start with (A.2)
$$d_QA_n^{\prime 67...75...51...10...0}=\sum [(d_Q\tilde
T_k^{...})T_{n-k}^{...}
\pm \tilde T_k^{...}d_QT_{n-k}^{...}].\eqno(C.4)$$
The upper indices of $\tilde T_k$ and $T_{n-k}$ on the r.h.s. are arbitrary
many indices
7,5,1,0 and at most one index 6. Consequently, we can insert the induction
hypothesis (C.2-3)
for $d_Q\tilde T_k$ and $d_QT_{n-k}$ and obtain (C.3) for the
$A'_n$-distributions and
similar for $R'_n,\,R''_n$. Therefore, we may define the normalization of
$R_n^{7...75...51...10...0}$ by (C.3). This procedure conserves (C.3) in the
splitting (A.7),
and the remaining steps do not destroy it either.

We turn to (C.2). The case $r=0$ is the assumption (2.50a). For $1\leq r\leq n$
we apply
$d_Q$ to (C.3) and use $(d_Q)^2=0$
$$d_QF_n^{7...75...51...10...0}=-i\sum_{j=r+t+1}^{r+t+s}(-1)^{(j-r-t)}
\d^jd_Q{\cal S}_rF_n^{67...75...51...151...10...0}-\eqno(C.5a)$$
$$-i(-1)^{s+1}\sum_{j=r+t+s+1}^n\d^jd_Q{\cal
S}_rF_n^{67...75...51...10...010...0}.\eqno(C.5b)$$
Next we insert again (C.3) into both terms on the r.h.s
$$(C.5a)=-i\sum_{j=r+t+1}^{r+t+s}(-1)^{(j-r-t)}\d^j\Bigl\{
i\sum_{l=r+t+1\>(l\not= j)}^{r+t+s}\pm\d^l{\cal
S}_rF_n^{67...75...51...151...151...10...0}+\eqno(C.6a)$$
$$+i(-1)^s\sum_{l=r+t+s+1}^n\d^l{\cal
S}_rF_n^{67...75...51...151...10...010...0}+\eqno(C.6b)$$
$$+F_n^{7...75...51...151...10...0}\Bigl\},\eqno(C.6c)$$
$$(C.5b)=-i(-1)^{s+1}\sum_{j=r+t+s+1}^n\d^j\Bigl\{
i\sum_{l=r+t+1}^{r+t+s}(-1)^{(l-r-t)}\d^l{\cal
S}_rF_n^{67...75...51...151...10...010...0}+\eqno(C.7a)$$
$$+i(-1)^{s+1}\d^j{\cal S}_rF_n^{67...75...51...10...050...0}+\eqno(C.7b)$$
$$+i(-1)^{s+1}\sum_{l=r+t+s+1\>(l\not= j)}^n\pm\d^l{\cal S}_r
F_n^{67...75...51...10...010...010...0}+\eqno(C.7c)$$
$$+F_n^{7...75...51...10...010...0}\Bigl\}.\eqno(C.7d)$$
(C.6b) and (C.7a) cancel. Similar to the reasoning after (2.50), the terms
(C.7b) and (C.7c)
vanish because of $F_n^{...5...\nu\mu}=-F_n^{...5...\mu\nu}$ and the different
signs of the
$(j,l)$- and the $(l,j)$-term in $\sum_{j,l\>(j\not=
l)}\pm\d^j\d^lF_n^{...010...010...}$.
The latter argument applies to (C.6a), too. (Due to (1.11) the $\pm$ in (C.6a)
is a factor
$(-1)^{(l-r-t)}$ if $l<j$, and a sign $(-1)^{(l-r-t-1)}$ for $l>j$.) It remains
$d_QF_n^{7...75...51...10...0}=$(C.6c)$+$(C.7d), which is the assertion
(C.2).$\quad\quad\quad\w$
\vskip 1cm
I would like to thank Ivo Schorn and Prof. G.Scharf for stimulating discussions
and A.Aste
for reading the manuscript. Finally, I thank my fianc\'ee
Annemarie Schneider for bearing with me during working at this paper.
\vskip 1cm
{\trm References}
\vskip 1cm
{\obeylines
[1] M.D\"utsch, T.Hurth, F.Krahe, G.Scharf, {\it N. Cimento A} {\bf 106}
(1993), 1029
[2] M.D\"utsch, T.Hurth, F.Krahe, G.Scharf, {\it N. Cimento A} {\bf 107}
(1994), 375
[3] M.D\"utsch, T.Hurth, G.Scharf, {\it N. Cimento A} {\bf 108} (1995), 679
[4] M.D\"utsch, T.Hurth, G.Scharf, {\it N. Cimento A} {\bf 108} (1995), 737
[5] M.D\"utsch, {\it N. Cimento A}, to appear
[6] H.Epstein, V.Glaser, {\it Ann. Inst. Poincar\'e A} {\bf 19} (1973), 211
[7] G.Scharf, ``Finite Quantum Electrodynamics'', 2nd. ed., Springer-Verlag
(1995)
[8] C.Becchi, A.Rouet, R.Stora, {\it Commun. Math. Phys.} {\bf 42} (1975), 127
    C.Becchi, A.Rouet, R.Stora, {\it Annals of Physics (N.Y.)} {\bf 98} (1976),
    287
[9] M.D\"utsch, preprint ZU-TH 30/95
[10] L.Baulieu, {\it Physics Reports} {\bf 129} (1985), 1
[11] F.Krahe, preprint DIAS-STP-95-01
[12] F.Krahe, preprint DIAS-STP-95-02
[13] T.Hurth, preprint ZU-TH 20/95, hep-th/9511139
[14] I.Schorn, preprints ZU-TH 96/16 and 96/17
[15] M.D\"utsch, F.Krahe, G.Scharf, {\it N. Cimento A} {\bf 106} (1993), 277
}
\bye